\documentclass[aps,pre,preprint,superscriptaddress]{revtex4-2}

\usepackage{graphicx}
\usepackage{amsmath}
\usepackage{color}
\usepackage{lineno}
\usepackage{amsfonts}
\usepackage{setspace}
\usepackage[T1]{fontenc}
\usepackage{siunitx}
\usepackage{hyperref}

\graphicspath{{figs/}}

\begin{document}

\title{Material elasticity determines scaling behaviour of cracking dynamics in porous materials: A precursor to crack percolation}

\author{Ruhul A. I. Haque}
\affiliation{Physics Department, St. Xavier's College, Kolkata 700016, India}
\affiliation{Condensed Matter Physics Research Centre, Jadavpur University, India}

\author{T. Dutta}
\affiliation{Physics Department, St. Xavier's College, Kolkata 700016, India}
\affiliation{Condensed Matter Physics Research Centre, Jadavpur University, India}
\email{tapati$\_$mithu@yahoo.com}


\begin{abstract}
The crack statistics of a 3-dimensional disordered porous material subjected to constant axial compressive loading have been investigated with regard to the elastic properties of the system.
 While cracking is a complex dynamics that involves material intrinsic properties like grain shape and size distribution, elastic properties of grain and cementing materials, and extrinsic properties of loading, in this work, the focus has been to check the dependence on the elastic properties of the bonding material. A 3-dimensional disordered system was constructed from spherical balls of varying radii that were chosen randomly from a log-normal distribution. The growth of micro-cracks with increasing compressive strain was monitored till the limit of the percolation crack. The two parameters varied were the bond stiffness constant and the bond strength of the material. Two distinct regimes of cracking rates were observed across a critical strain $\epsilon_{knee}$ that manifested as a knee in the cumulative crack-strain plot. The critical strain $\epsilon_{knee}$ and the strain at the percolation point $\epsilon_{perc}$ showed a power law dependence on the elastic property of the bond material. Individual micro-cracks were observed to grow sharply to a maximum value $N^{k_{b}}_{max}$, after which the number of new micro-cracks decreased, showing a long tail. The maximum $N^{k_{b}}_{max}$ was found to correspond to the strain $\epsilon_{knee}$, thus indicating that pre-$N^{k_{b}}_{max}$ cracking brittle, followed by ductile cracking behaviour of system. Lastly, we show that there exists a robust relation between $\epsilon_{knee}$ and $\epsilon_{perc}$ that is a power-law where the exponent is a function of the material elastic property. As  $\epsilon_{knee}$ can be determined from acoustic signals associated with micro-cracks, our proposed relation can act as a warning towards critical strain resulting in crack percolation.
\end{abstract}


\maketitle


\section{Introduction}
Understanding fracture processes and their correlation to material properties continues to be an active area of research for both engineers and scientists since the introduction of Griffith's theory \cite{Griffith1921} that provided the criteria that materials cracked open to create new interfaces as a means to release accumulated compliant or tensile stress. Since then, much experimental and theoretical research has indicated that cracks can appear as disjoint failures that can join, resulting in crack avalanches of different sizes and patterns that depend on intrinsic material properties and external forcing modes \cite{Wiley2015, Biswas2015}. Fractures of porous systems like granular solids, rocks, and colloidal systems that constitute a large part of our daily life are highly dependent on pressure, grain size and shape distribution, micro-geometry of the pore space \cite{Jaeger1976, Hansen2010, Herrmann2010} and show a complexity very different from the fracture in ordered metallic systems \cite{Bouchaud2001}. Experimental research with porous systems indicates that failure starts with the nucleation of micro-cracks that are manifested as bursts of acoustic energy \cite{Martin1997, Eberhardt1997}. The microcracks grow, merge and finally result in the macro-failure of the system \cite{Lockner1991, Benson2008}. Failure modes can vary from axial splitting to shear band formation manifested as a non-linear transition from brittle to ductile cracking\cite{Kaiser2008, Paterson2004}. Large-scale simulation studies using lattice models \cite{Herrmann2014, Tzschichholz1995, Zapperi1997, Caldarelli1996} and statistical analysis of results have helped to provide a good description of the geometrical and topological aspects of fracture and scaling concepts \cite{Hansen1994}. The complete breakdown of a system is reported to be preceded by avalanches of micro-cracks \cite{Baro2018, Diksha2023} reported in different types of materials - from wood \cite{Garcimartin1997}, to glass \cite{Maes1998}, from volcanic activity \cite{Diodati1991} to earthquakes  \cite{Beno1956}. Some of the quasi-static lattice models commonly used to model fracture include the random fuse model \cite{Arcangelis1985, Kahng1988} and the spring-network model \cite{Ray1996, Sadhukhan2019}; however, these are limited by the fact
that deformable contacts, slips, and rotations, as well as the presence of discontinuities, cannot be explicitly captured herein.

In this work, we study fracture statistics and its dependence on material properties of a porous system subjected to axial compression using the Distinct Element Method (DEM)\cite{Lisjak2014, Potyondy2004}.  Spherical grains of different sizes are dropped under gravity to form a 3-dimensional structure. The radii of the spheres were chosen randomly from a log-normal probability distribution. The system is then subjected to compressive loading, applied quasi-statically.
 One can picture the compression as the effect of a rigid wall in contact with the upper surface and descending with a velocity $V_{w}$. We assumed that the grains could deform under compressive stress. Thus, the contacts varied from points to planes. The grains are assumed to be cemented at their contacts, and the cementation material has elastic properties that are different from those of the grains. We monitor the loading effects via calculation of stress and strain accumulation, the number of micro-cracks created up to crack avalanche, after which the system develops one or more system-spanning percolating cracks. In this work, we explore the dynamic cracking process as a function of the elastic modulus and breaking threshold of cementing material between the grains. The other factors of grain size distribution, elastic property of grain, and wall velocity are kept constant. The applied macroscopic load is carried by the grain and cement skeleton in the form of force chains that propagate from one grain to the next across grain contacts. DEM is used to calculate the forces that can be compressive, tensile, or shear in nature.

On analysis of the \textit{cumulative crack statistics}, we observed two different regimes of cracking - (i) a zone of micro-crack creation,  followed by (ii) a zone of micro-crack merging that leads to final percolating cracks. For a constant quasi-static loading, the transition from one zone to the other occurs across a critical value of the variable elastic property of the cementing material. Interestingly, we observed that both the critical value of changeover from one zone of cracking to the other and the point of total percolation showed a scaling behaviour with the elastic property of the cementing material, albeit with different exponents. Examination of individual microcrack bursts revealed that the largest burst of microcracks also followed a power law behaviour with the breaking threshold of the cementing material. The scaling behaviour of load distribution with cracking for fixed material properties has been reported earlier \cite{Bodaballa2024}. Ferenc et al. \cite{Kun1999, Kun2014}  have shown that fragment mass follows a scaling behaviour with a dimensionless energy measure that is a function of material properties. 

In the following sections, we shall describe the methodology of structure generation and the DEM. This will be followed by the sections describing our results and the corresponding discussion, and finally, the conclusion of our work.

\section{Structure generation and DEM} 
The disordered porous system was generated by depositing spherical particles of elastic modulus $Y$ under gravity in a 3-dimensional box of $1 \times 1\times 1 $ \si{\centi\metre\cubed} size. The radii $R$ of the spherical particles were chosen randomly from a log-normal probability distribution. 
It is assumed that:
\begin{itemize}
\item The particles can both translate and rotate independently of each other.
\item Two particles $A$ and $B$ are said to be in contact if the distance $d$ between their centres  satisfies the condition $d \leq R_{A} + R_{B}$.
\item Particles interact via contact points only, where a contact comprises only two particles.
\item Particles are allowed to overlap over a small region at the point of contact; however, the overlaps are small in relation to particle size.
\item Bonds of finite stiffness and breaking thresholds exist at contacts. These bonds can break if the threshold is crossed.
\item Newton's second law is used to determine the translational and rotational motion of each particle, while the force-displacement law is used to update the contact forces arising from the relative motion at each contact.
\item Dynamics is implemented by updating particle position and bond states in a time step $\Delta t$ small enough to assume constant velocity and acceleration values.
\item $\Delta t$ is chosen in a manner such that disturbances due to a particle cannot propagate further than its nearest neighbours.
\end{itemize}

\begin{figure}
\includegraphics[width=\linewidth]{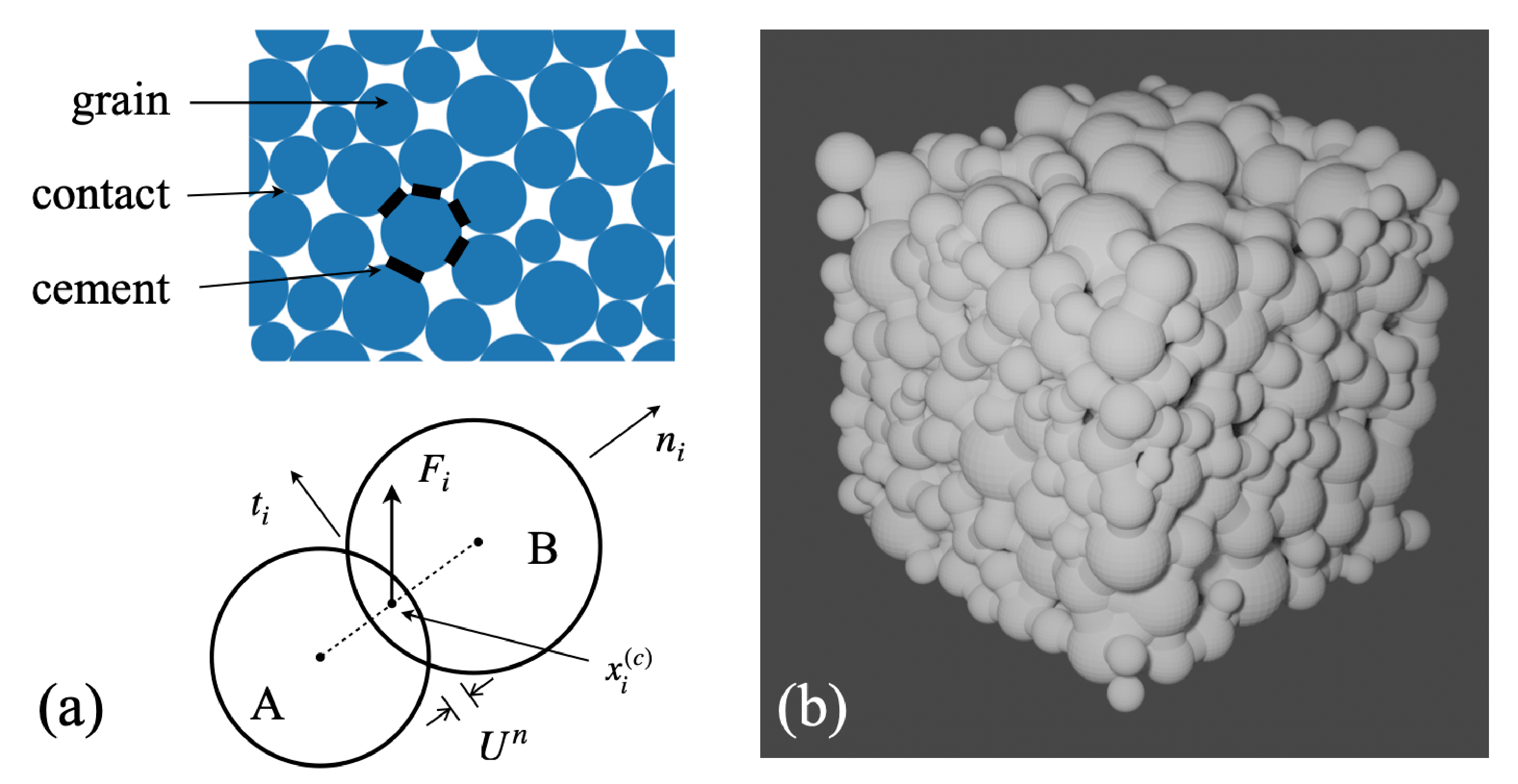}
\caption{(a) Schematic diagram of forces acting on particles in DEM. (b)3-dimensional rectangular parallelopiped constructed from spheres having a log-normal distribution of mean radius $0.04$ \si{\centi\metre} and deviation $0.003$ \si{\centi\metre}.}
\label{structure}
\end{figure} 

 The interaction between the particles is treated as a dynamic process with equilibrium states occurring whenever internal forces become zero. The application of an external force on the top face of the structure causes a disturbance by particle movements relative to each other, with the speed of propagation being dependent on particle contact distribution and material properties. The force-displacement behaviour at each contact is described by the normal and shear stiffness, $k_{n}$ and $k_{s}$ and the coefficient of friction $\mu$. The calculation of the net normal force $F^{n}$ and the net shear force $F^{s}$ following the scheme described by Potyondy et al. \cite{Potyondy2004}.

The contact force, $\mathbf{F_{i}}$ of, say, particle $A$ on $B$ is resolved into its normal and tangential components with respect to the contact plane as
  \begin{equation}
     \mathbf{ F_i }= F^{n}\mathbf{n_i} + F^{s}\mathbf{t_i}
  \end{equation}
where $F^{n}$ and $F^{s}$ are the normal and shear components respectively with $\mathbf{n_i}$ and $\mathbf{t_i}$ their corresponding direction vectors that define the contact plane.  The normal force is thus
\begin{equation}
F^{n} = K^{n}U^{n}
\end{equation}
where $U^{n}$ is the overlap between two spheres, Fig. \ref{structure}(a), and $ K^{n}$ is the normal stiffness constant of the particles. The change in shear force $\Delta F^{s}$ for a relative shear displacement $\Delta U^{s}$ is given by
 \begin{equation}
\Delta F^{s} = - K^{s}\Delta U^{s}
\end{equation}
where $K^{s}$ is the particles' shear stiffness.
At any time step of interval $\Delta t$, the relative displacement is given by
 \begin{equation}
\Delta U_{i} = V_{i}\Delta t
\end{equation}
where $V_{i}$ is the contact velocity. The contact velocity $V_{i}$ is a function of both the translational and rotational velocities of the $i^{th}$ particle. The relative shear displacement can be similarly written as 
\begin{equation}
\Delta U_{s} = (V_{i}- V^{n}_{i})\Delta t
\end{equation}
If there exists a gap between the particles, i.e. $U^{n}\leq 0$, then both the normal and shear forces are $0$.

The cement-based bonds between the particles are envisioned as elastic springs uniformly distributed over a circular cross-sectional area that can transmit both force $F_{i}$ and moment $M_i$ between the particles. Each bond has normal and shear stiffness per unit area, $k_n$ and $k_s$, respectively; tensile and shear strengths, $\sigma_c$ and $\tau_c$. The radius of a parallel bond is determined by
\begin{equation}
\bar{R} = \lambda \text{ min}(R^{A},R^{B})
\end{equation}
where $\lambda$ is the bond-radius multiplier. We kept the value of $\lambda$ constant throughout the simulation with $\lambda = 1.0$.

 As before, the total force and moment of each bond can be resolved into normal and shear components with respect to the contact plane. The initial forces and moments on the bonds are set to zero. Subsequent relative displacements and rotations increase the elastic force and moment values that are added onto the bonds and are given by 
 \begin{eqnarray}
 \Delta \bar{F}^{n} = k^{n}A\Delta U^{n}\\
 \Delta \bar{F}^{s} = - k^{s}A\Delta U^{s}\\
 \Delta \bar{M}^{n} = -k^{s}J\Delta\theta^{n}\\
 \Delta \bar{M}^{s} = -k^{n}I\Delta\theta^{s}
 \end{eqnarray}
 where $\Delta\theta^{n}$and $\Delta\theta^{s}$ are rotational increments in the normal and tangential directions; $A, I$ and $J$ represent the cross-sectional area, moment of inertia and the polar moment of inertia of the bonds, respectively. 
 The maximum stress on a bond can be calculated from the beam theory --
  \begin{eqnarray}
      \sigma^{max} = \frac{-\bar{F}^{n}}{A} + \frac{|\bar{M}^{s}|\bar{R}}{I} \\
      \tau^{max} = \frac{|\bar{F}^{s}|}{A} + \frac{|\bar{M}^{n}|\bar{R}}{J}
   \end{eqnarray}
 If the maximum tensile stress exceeds the tensile strength ($\sigma^{max}\geq \sigma_c$) or the maximum shear stress exceeds the shear strength ($\tau^{max}\geq \tau_c$), the spring breaks. The ratios $(K^{n}/K^{s})$ and $(k^{n}/k^{s})$ are related to the Poisson ratio of the material. Increasing these ratios for fixed grain shape and packing increases the Poisson ratio.

\section{Results and Discussion} 
Spherical particles were dropped under gravity and allowed to settle using DEM to construct 3-dimensional porous granular media. The system is subjected to an axial compressive stress exerted on the top surface of the rectangular parallelepiped. One can assume a rigid wall in contact with the top surface moving down with a velocity $V_{w}$. It is assumed that all spheres constituting the system can break free if all bonds connecting it to other spheres fail. Under the DEM scheme, the bonds connecting the spheres can deform through elongation and/or twist. Once a bond breaks, the load is distributed amongst the other intact bonds. Thus, as more micro-cracks develop, i.e., bonds break, the stress on the system increases. Although crack formation can be affected by all the intrinsic characteristics of grain and bonding material and the external loading, we only vary the cementing bond properties in this study.  Thus, the grain (sphere) elastic properties are kept constant in this work. As the normal and shear stiffness are related via the Poisson ratio of the material, they can be expressed as
\begin{equation}
    k^{n} = \nu k^{s} = k_{b}
\end{equation}
where $\nu$ is a constant with a value taken as $2.5$. The shear strength value is taken the same as the normal strength, $\sigma_{c} = \tau_{c}$.
Our studies were done on a disordered system having a log-normal distribution of radii with mean radius $0.04$ \si{\centi\metre} and a deviation of $0.003$ \si{\centi\metre}, Fig. \ref{structure}(b).
We investigated crack statistics by (i) varying the stiffness constant $k_b$ of the bond spring, keeping the bond spring threshold $\sigma_c$ constant; (ii) varying the spring threshold $\sigma_c$  of the bond while keeping the spring stiffness $k_b$ constant. The wall velocity is kept constant at $1$ \si{\centi\metre\per\second}. All results are averaged over several configurations to obtain convergence in results.

\subsection{Fixed bond strength, varying stiffness constant}

\begin{figure}
\includegraphics[width=\textwidth]{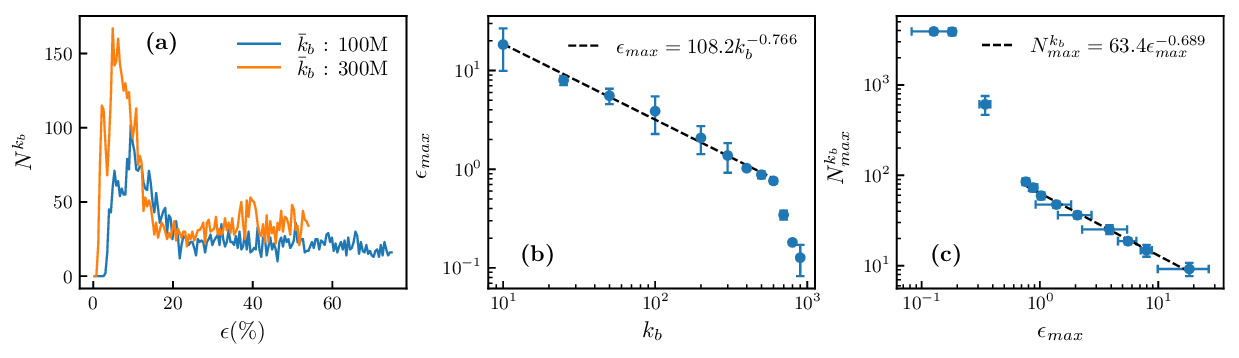}
\caption{(a)Variation of the number of cracks $N^{k_b}$ versus axial strain for different bond stiffness and constant bond strength $\sigma_c$. (b) Log-Log plot of strain at maximum crack number $N^{k_{b}}_{max}$ versus bond stiffness $k_{b}$.(c) Log-Log plot of strain at maximum crack number $N^{k_{b}}_{max}$ versus corresponding strain. }
\label{max_cracks_constant_tau100}
\end{figure}

Under the condition of fixed bond strength $\sigma_{c}= 100$ \si{\mega\pascal} and different values of stiffness $k_{b}$, the number of micro-cracks was plotted with increasing axial strain as shown in Fig. \ref{max_cracks_constant_tau100}(a). The number of micro-cracks $N^{k_b}$ sharply increased with an increase in strain up to a maximum peak value $N^{k_{b}}_{max}$, after which the crack number decreased slowly, showing a long tail. $N^{k_{b}}_{max}$ was found to decrease with decreasing $k_{b}$, but had longer tails. Higher bond stiffness of the springs representing the cementing material implies that many springs in the system can accumulate strain for a greater number of time steps before the breaking threshold is reached. The crack burst sizes are bigger, contributing to the higher value of $N^{k_{b}}_{max}$ observed in this case. The number of intact springs that survive after the maximum is smaller. These, too, survive a short time due to accumulated strain, so a shorter tail is observed here. It is observed that the strain value $\epsilon_{max}$ corresponding to $N^{k_{b}}_{max}$ decreased with increasing bond stiffness $k_{b}$ as expected. Plotting the variation of both $\epsilon_{max}$ and $N^{k_{b}}_{max}$ with bond stiffness $k_{b}$ showed a power-law dependence in each case as shown in Figs. \ref{max_cracks_constant_tau100}(b) and (c).

The asymmetric distribution of the micro-cracks may be explained in the following way: with an increase in axial strain, the initial fracturing is more brittle, as manifested in the sharp growth rate of cracks. However, the cracks that begin as small bursts rapidly spread through the system as more and more load is shared by the intact bonds to compensate for the loss of broken bonds.  Beyond a critical strain $\epsilon_{max}$, the micro-cracks start merging, and the system moves towards a percolating failure with fewer bonds getting broken with a further increase in strain. 
For our log-normal particle size distribution, the graphs in Figs. \ref{max_cracks_constant_tau100}(b) and (c) indicate that the strain value at the maximum number of micro-cracks follows a power-law trend with a negative exponent of $0.766$ as the bond stiffness increases. However at very large values of bond stiffness -- of the $k_{b}\geq 160$ -- the value of  $\epsilon_{max}$ falls rapidly. This can be explained by the fact that a very high bond stiffness allows a quick accumulation of stress as the bond cannot relax. Hence the breaking threshold of the bonds is achieved relatively quickly and the bonds break leading to system failure. This is also reflected in a very high value of $N^{k_{b}}_{max}$ at $k_{b}\geq 160$ as observed in Fig. \ref{max_cracks_constant_tau100} (c). Additionally, the log-log plot of the maximum number of micro-cracks $N^{k_{b}}_{max}$ versus the corresponding strain value for different $k_{b}$ also demonstrates a power-law relationship with a negative exponent of $0.689$.

 \begin{figure}
\includegraphics[width=\textwidth]{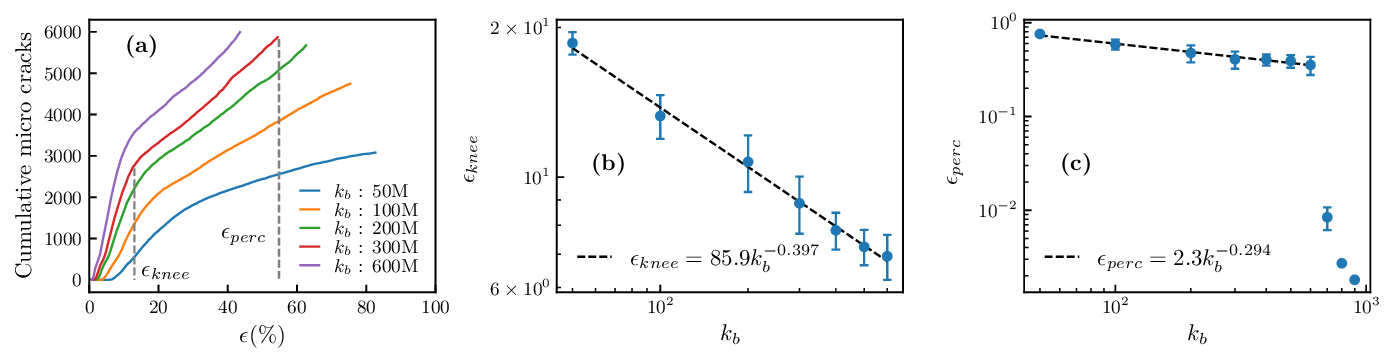}
\caption{Crack statistics for constant spring threshold $\sigma_c= 100$ \si{\mega\pascal}, and variable bond stiffness constants $k_{b}= 50, 100, 200, 300, 600$. Values are in units of \si{\mega\pascal\per\metre}. (a) Cumulative micro-cracks versus strain percentage. The dotted line is indicative of the transition between different cracking modes for $k_{b}= 300$ \si{\mega\pascal\per\metre}. (b)Variation of $\epsilon_{knee}$ with $k_{b}$ follows a power law with exponent $m_{knee} = - 0.397$ (c) Variation of $\epsilon_{perc}$ with $k_{b}$ follows a power law with exponent $m_{perc} = - 0.294$}
\label{Constant_tau100}
\end{figure} 
  
  Figure \ref{Constant_tau100}(a) shows that when the spring stiffness $k_{b}$ is varied, keeping the bond breaking threshold $\sigma_c = 100$ \si{\mega\pascal} constant, the cumulative cracks in the system increase with the percentage of strain in the system. Every curve shows two distinctly different growth regions about a knee point, the strain at the knee being denoted by $\epsilon_{knee}$. Interestingly, a comparison of Figs.\ref{max_cracks_constant_tau100}(a) and \ref{Constant_tau100}(a) show that the maximum number of micro-cracks corresponds to the knee point strain $\epsilon_{knee}$. The endpoint of each curve corresponds to the point of percolation in the system; the corresponding strain is denoted as $\epsilon_{perc}$. Both $\epsilon_{knee}$ and $\epsilon_{perc}$ decrease with increasing values of $k_{b}$, and show a power-law dependence, Figs. \ref{Constant_tau100} (b) and (c), of the form:
\begin{equation}
\epsilon_{knee} = Ak_{b}^{-m_{knee}}
\label{knee}
\end{equation}  
and
\begin{equation}
\epsilon_{perc} = Bk_{b}^{-m_{perc}}
\label{perco}
\end{equation}
The exponents have values $m_{knee} = 0.397$  and $m_{perc} = 0.294$; $A$ and $B$ are constants characteristic of the system. In Fig. \ref{Constant_tau100}(c), we see that the system fails at $k_{b}\geq 160$ for the reasons discussed earlier. Combining Eqs. (\ref{knee}) and (\ref{perco}), we get a relation between $\epsilon_{knee}$, $\epsilon_{perc}$, $m_{knee}$ and $m_{perc}$ given by  
\begin{equation}
\epsilon_{perc} = D{\epsilon_{knee}}^m
\label{combo}
\end{equation}
 Thus Eq. (\ref{combo}) indicate that $\epsilon_{knee}$ has a power-law scaling with $\epsilon_{perc}$ with the exponent $m = \frac{m_{perc}}{m_{knee}} = 0.740$
\begin{figure}[h]
  \includegraphics[width=\linewidth]{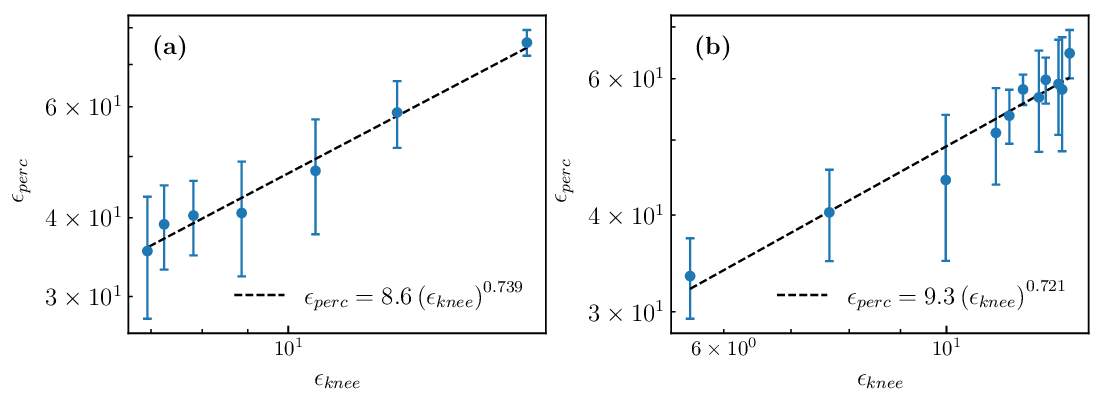}
  \caption{Log-log plot of $\epsilon_{perc}$ versus $\epsilon_{knee}$ (a) For constant breaking threshold $\sigma_{c}= 100$ \si{\mega\pascal} and varying stiffness constant $k_{b}$. A power-law behaviour is indicated with exponent  $m=0.739$. (b) For constant stiffness constant $k_{b} = 400$ \si{\mega\pascal\per\metre} and varying breaking threshold $\sigma_c$. A power-law behaviour is indicated with exponent  $m=0.721$}
  \label{verify}
\end{figure}

Using the data points from our simulation, we construct the variation of $\epsilon_{perc}$ versus $\epsilon_{knee}$ as shown in Fig. \ref{verify}(a).  A power-law behaviour is observed with the exponent $m = 0.739$  agreeing almost exactly with the theoretical value of $0.740$. We repeated this study with two other constant bond strength values of $\sigma_c = 500$ \si{\mega\pascal} and $1000$ \si{\mega\pascal}, and variable bond stiffness constants. The nature of the cumulative cracks versus strain graphs showed similar crack statistics as observed for $\sigma_c = 100$ \si{\mega\pascal}.

\subsection{Fixed stiffness constant, varying bond strength}
\begin{figure}
\includegraphics[width=\textwidth]{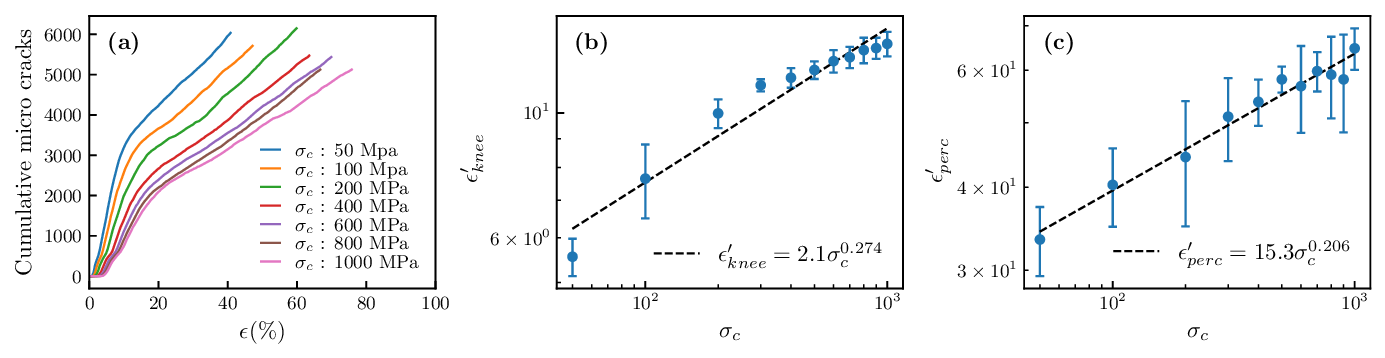}
\caption{Crack statistics for constant spring stiffness $k_{b} = 400$ \si{\mega\pascal}, and variable spring breaking thresholds $ \sigma_{c} = 50, 100, 200, 400, 600, 800, 1000 $. Values are in units of \si{\mega\pascal}. (a) Cumulative micro-cracks versus strain percentage. (b)Variation of $\epsilon^{\prime}_{knee}$ with $\sigma_{c}$ follows a power law with exponent $m^{\prime}_{knee} =  0.274$ (c) Variation of $\epsilon^{\prime}_{perc}$ with  $\sigma_{c}$ follows a power law with exponent $m^{\prime}_{perc} = 0.206$}
\label{Constant k400}
\end{figure}

In the situation where crack statistics on the same system was studied for a fixed bond stiffness constant $k_{b}=400$ \si{\mega\pascal\per\metre} and the breaking threshold of spring $\sigma_c$ varying from \SI{50}{\mega\pascal} to \SI{1000}{\mega\pascal}, the variation of cumulative micro-cracks increased with axial strain as expected. Similar to the situation where $\sigma_c$ was fixed and $k_{b}$ varied, two distinctly different growth rates were observed across a knee strain, Fig. \ref{Constant k400}(a). The strain at the knee point $\epsilon^{\prime}_{knee}$ and the percolation point $\epsilon^{\prime}_{perc}$ for each plot of bond strength was noted. These values, when plotted versus bond strength on a log-log scale, could be fitted approximately by a straight line indicative of a scaling behaviour, Figs. \ref{Constant k400}(b) and (c). The exponents  $m^{\prime}_{perc}$ and $m^{\prime}_{knee}$ had values $0.206$ and $0.274$ respectively. 

 Using the same procedure as described between Eqs. (\ref{knee} -- \ref{combo}),  a relation between $\epsilon^{\prime}_{knee}$, $\epsilon^{\prime}_{perc}$,  $m^{\prime}_{perc}$ and $m^{\prime}_{knee}$ was constructed :
\begin{equation}
\epsilon^{\prime}_{perc} = D^{\prime}{\epsilon^{\prime}_{knee}}^{m^\prime}
\label{comboprime}
\end{equation}
 Thus Eq. (\ref{comboprime}) is similar to Eq. (\ref{combo}), i.e., $\epsilon^{\prime}_{knee}$ has a power-law scaling with $\epsilon^{\prime}_{perc}$ with the exponent $m^{\prime} = 0.752$. 
 Using the data points of our simulation, we constructed the variation of $\epsilon^{\prime}_{perc}$ versus $\epsilon^{\prime}_{knee}$ on a log-log scale as shown in Fig. \ref{verify}(b) for constant bond strength $k_{b}=400$ \si{\mega\pascal\per\metre}. A power-law behaviour between the parameters was obtained with the exponent $m^{\prime} = 0.721$, which is close to the theoretically predicted value of  $0.752$. 
 
 It may be recalled that the systems under examination are highly disordered. The size distribution of the grains adds to the complexity, and the Distinct Element Method (DEM) can be computationally expensive when applied to 3-dimensional systems. The difference between the theoretical and experimental values of the exponents $m$ and $m^\prime$ can be better matched when crack statistics are calculated on a larger system.

From our results and analysis so far, we propose that the scaling relation between the strain at the knee $\epsilon_{knee}$ and at the percolating point $\epsilon_{perc}$ as given by Eqs. (\ref{combo}) and (\ref{comboprime}), can have useful applications as an indicative precursor to the percolating point of a 3-dimensional system. Micro-cracks are detected as acoustic signals in experiments. As DEM factors in individual particle interactions and resultant deformations in relation to the rest of the system, with all system parameters known, the exponent $m$/$m^{\prime}$ for a porous system can be estimated accurately. One can detect $\epsilon_{knee}$ via acoustic emissions and can have an estimate of $\epsilon_{perc}$. Though a percolation plane ultimately leads to complete failure in a 3-dimensional structure, the pre-knowledge of $\epsilon_{perc}$ certainly shall act as a red flag.

\section{Conclusions}
In this work, we studied the effect of material properties on the crack statistics of a porous granular system under compressive strain. For this purpose, a 3-dimensional disordered porous system was generated by dropping spherical balls with different radii whose values were chosen from a suitable log-normal distribution. The rigid balls (grains) are assumed to be bonded with a binding cement, represented by Hookean springs, having elastic properties different from that of grains. A rigid wall in contact with the upper surface of the rectangular parallelepiped was allowed to descend with a constant velocity, thereby exerting a compressive axial strain on the system. The stress-strain progression and cracking dynamics were monitored in a quasi-static manner, and data was analyzed to check the role of elastic properties of the binding cement on micro-cracks to the final percolating crack in the system.

The system cracking statistics were examined under two scenarios - one where the bond spring stiffness was varied for a fixed bond strength and another case where the bond strength was varied, keeping the spring stiffness constant. All other system properties were fixed and kept identical for these two scenarios. In both situations, the cumulative crack count increased with increasing strain in the system, as expected. However, the rate of increase changed drastically at a critical strain value $\epsilon_{knee}$, the value which decreased with higher values of bond stiffness and increased with increasing bond strength. The system strain at the point of percolation $\epsilon_{perc}$ also changes similarly with the elastic property and the strength of the bonding material. In both the scenarios studied by us, $\epsilon_{knee}$ and $\epsilon_{perc}$ showed a power-law behaviour with the corresponding variable of the bonding cement, with the exponents different for each scenario. From our simulation, we constructed a scaling relationship between $\epsilon_{knee}$ and $\epsilon_{perc}$ with the exponent being a function of the material property (stiffness or strength) of the bonds. When we tested our crack statistics for different cementing properties, the proposed relationship was found to be robust.

The robustness of the relationship, as provided in Eqs. (\ref{combo}) and (\ref{comboprime}) leads us to propose that one can anticipate the onset of percolation cracks in a system. Obtaining statistics of micro-cracks via their associated acoustic signals is a standard practice adopted by crack experimentalists. As standard values of stiffness and strength modulus of common cementing material are generally obtainable, and exponent values are known from simulation studies, one can predict the limit of strain that a system can sustain before crack percolation occurs. While it is true that a 3-dimensional system needs to have a percolating plane to fragment it, the percolation limit being a certain damage line can act as a red flag before total system failure.

\begin{acknowledgments}
R A I Haque acknowledges support from UGC-funded research fellowships (UGC-ref No. 1435/ CSIR-UGC NET JUNE 2017). R.A.I. Haque acknowledges S. Roy for useful discussions on DEM.
\end{acknowledgments}

\bibliography{ref_frac}

\end{document}